\documentclass[aps,prl,reprint,showpacs,superscriptaddress]{revtex4-1} 
\usepackage{hyperref} 
\hypersetup{breaklinks=true}
\usepackage{amssymb}
\usepackage{amsmath}
\usepackage{mathtools}
\usepackage{lineno}
\usepackage{braket}
\usepackage{hhline}
\usepackage{graphicx}
\usepackage{dcolumn}
\usepackage{bm}
\usepackage[usenames, dvipsnames]{xcolor}
\usepackage{setspace}
\usepackage[caption=false]{subfig}

\begin{document}

\title{Hybrid quantum computation gate with trapped ion system}
\author{H.~C.~J.~Gan}
\author{Gleb Maslennikov}
\author{Ko-Wei Tseng}
\author{Chihuan Nguyen}
\affiliation{Centre for Quantum Technologies, National University of Singapore, 3 Science Dr 2, 117543, Singapore}
\author{Dzmitry Matsukevich}
\affiliation{Centre for Quantum Technologies, National University of Singapore, 3 Science Dr 2, 117543, Singapore}
\affiliation{Department of Physics, National University of Singapore, 2 Science Dr 3, 117551, Singapore}

\begin{abstract}
 The hybrid approach to quantum computation simultaneously utilizes both discrete and continuous variables which offers the advantage of higher density encoding and processing powers for the same physical resources. Trapped ions, with discrete internal states and motional modes which can be described by continuous variables in an infinite dimensional Hilbert space, offer a natural platform for this approach. A nonlinear gate for universal quantum computing can be implemented with the conditional beam splitter Hamiltonian $\ket{e}\bra{e}(\hat{a}^{\dagger}\hat{b} + \hat{a}\hat{b}^{\dagger})$ that swaps the quantum states of two motional modes, depending on the ion's internal state. We realize such a gate and demonstrate its applications for quantum state overlap measurements, single-shot parity measurement, and generation of NOON states. 
\end{abstract}


\date{\today}
\maketitle

While quantum computers are expected to provide solutions to computational problems deemed classically intractable today, the issue of scability remains a challenge~\cite{Brown2016,Bruzewicz2019}. The estimated number of logical qubits required to perform complex and useful computations range from hundreds to thousands, exceeding present day capabilities~\cite{Monz2016,Debnath2016,Linke2017,Linke2018}. The overhead from implementing error correction codes steepens the requirement further, to thousands and millions~\cite{Fowler2012,Reiher2017} of physical qubits. 

An alternative means of obtaining a larger Hilbert space can come from considering continuous variables~\cite{Lloyd1999,NielsenChuang2010,HKLau2016}, which in the case of trapped ions, can be realized in motional modes. Naturally offering a large Hilbert space per ion, the use of motional modes allows the continuous variable quantum computation (CVQC)~\cite{GKP2001,Braunstein2005,Shen2018} to be adapted to trapped ion systems~\cite{Ortiz2017} with well established experimental techniques~\cite{Meekhof1996,Leibfried1996,Olmschenk2007,Fluhmann2019}. A universal set of CVQC gates include Gaussian gates such as displacement, squeezers, beam splitter and rotations, and a non-Gaussian gate~\cite{Lloyd1999}. The latter can be achieved in a system of trapped ions via the nonlinear interactions between motional modes~\cite{Roos2008,Ding2017a,Ding2017b,Ding2018}. 

Recently, the hybrid approach where both discrete and continuous variables are utilized simultaneously has garnered attention due to its potential to overcome intrinsic limitations of each approach~\cite{Lloyd2003,Andersen2015}. In the context of this approach, a non-Gaussian gate could also be implemented via a nonlinear interaction that arises from the coupling between internal and motional degrees of freedom of trapped ions~\cite{HKLau2016}. 

Here we report an experimental realization of a conditional beam splitter (CBS) gate with a single trapped $^{171}\mathrm{Yb}^+$ ion where, conditioned on the spin state of the ion, two motional modes undergo the beam splitter transformation~\cite{MandelWolf1995}. We show that the CBS gate is a viable candidate as a non-Gaussian gate for CVQC, and demonstrate algorithms utilizing this gate.
The interaction Hamiltonian describing the operation is
\begin{equation}
\label{eq:CBS_H}
\hat{H}_\textrm{CBS} =  \hbar \xi \ket{e}\bra{e}( \hat{a}^{\dagger} \hat{b}\operatorname{e}^{i\upsilon} + \hat{a} \hat{b}^{\dagger}\operatorname{e}^{-i \upsilon}),
\end{equation}
where $\hat{a} \,(\hat{a}^{\dagger})$ and $\hat{b} \,(\hat{b}^{\dagger})$ are the annihilation (creation) operators of the motional modes $a$ and $b$, $\xi$ is the coupling strength, and $\upsilon$ is a phase factor set by system parameters. A phase $\upsilon=0$ gives rise to the unitary transformation 
$\hat{U}(t)=\mathrm{exp}(-\mathrm{i} \xi t \ket{e}\bra{e}( \hat{a}^{\dagger} \hat{b} + \hat{a} \hat{b}^{\dagger}))$. The CBS gate is realized by applying the coupling for a time of $\tau= \frac{\pi}{2\xi}$, $\hat{U}(\tau)=\hat{U}_\textrm{CBS}$. When applied to Fock states, we obtain the transformation
\begin{align}
\label{eq:CBS_U}
\hat{U}_\textrm{CBS}\ket{g,n,m} = & \ket{g,n,m}, \nonumber \\
\hat{U}_\textrm{CBS}\ket{e,n,m} = & (-i)^{n + m}\ket{e,m,n},
\end{align}
where $\ket{g}\,(\ket{e})$ represents the ground (excited) spin state of the ion, and $\ket{n}$, $\ket{m}$ are the Fock states of motional modes $a$ and $b$.

The experiments are carried out on a single trapped $^{171}\mathrm{Yb}^+$ ion confined in a standard linear RF Paul trap, with secular trap frequencies $( \omega_\textrm{x} , \omega_\textrm{y} , \omega_\textrm{z}) = 2\pi\times(0.96 , 1.31, 0.53)$\,MHz (Fig.~\ref{fig:exp_setup}). The experimental sequences for all results reported in the paper begins with 2\,ms of Doppler cooling, followed by 4\,ms of Sisyphus cooling~\cite{Ejtemaee2017}. The x- and y-direction radial motional modes (corresponding to modes $a$ and $b$ respectively) are further sideband cooled down to their ground state, by driving two-photon Raman transitions via the hyperfine states $\ket{^2S_{1/2},F=0, m_F=0}$ and $\ket{^2S_{1/2},F=1, m_F=0}$. The average phonon numbers after sideband cooling are $\bar{n}_a = 0.004(3)$ and $\bar{n}_b=0.011(3)$. The heating rates are $\dot{\bar{n}}_a=19.9(9)\,s^{-1}$ and $\dot{\bar{n}}_b=44(3)\,s^{-1}$. A frequency-doubled mode-locked picosecond Ti:Sapphire laser, with central frequency of 374\,nm and repetition rate of 76.2\,MHz, generates the pair of beams responsible for the Raman transition~\cite{Hayes2010,Ding2017a,Ding2017b}. Detection of motional state is done by coupling it to the internal spin state via driving the blue or red sidebands~\cite{Meekhof1996,Leibfried1996,Leibfried2003}. The spin state is then detected with standard fluorescence techniques~\cite{Olmschenk2007}.

With linear and mutually orthogonal polarization of the Raman beams (Fig.~\ref{fig:exp_setup}a), a state dependent optical dipole force can be achieved~\cite{Wineland1998,Ding2014}. Interfering the Raman beams gives rise to a running optical lattice where polarization oscillates between left and right circular, at a frequency $\omega_L$. This causes the state $\ket{^2S_{1/2},F=1,m_F=1}=\ket{e}$ to experience a modulated ac Stark shift, whereas $\ket{^2S_{1/2},F=0, m_F=0}=\ket{g}$ does not (Fig.~\ref{fig:exp_setup}d). When $\omega_L$ = $|\omega_x - \omega_y|$, the Hamiltonian Eq. (\ref{eq:CBS_H}) is realized, and for $\tau\approx400\mu\,$s the transformation Eq.~(\ref{eq:CBS_U}) is obtained.
\begin{figure}
\centering
\includegraphics[width=0.85\columnwidth]{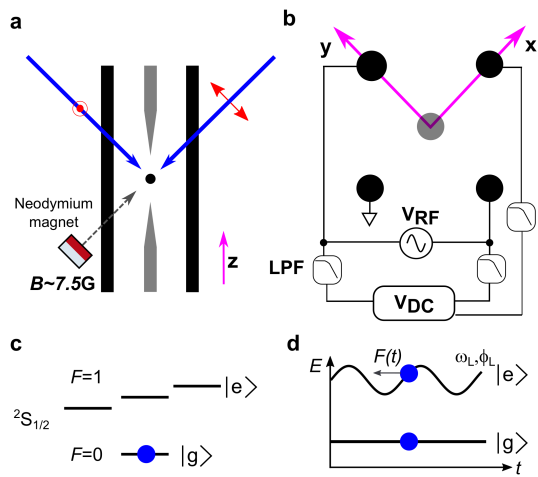}
\caption{
\label{fig:exp_setup}
(a) Experimental setup. A linear rf-Paul trap confines a single $^{171}\mathrm{Yb}^+$. Raman beams represented by blue arrows, with polarization denoted in red, form a running optical lattice on the ion. The beat note frequency $\omega_L$ and phase $\phi_L$ of the lattice are controlled by rf signals sent to acousto-optical modulators in each Raman beam (not shown).
(b) Simplified schematic of the RF Paul trap. In addition to V$_\textrm{RF}$, DC voltages are added to increase the difference between radial trap frequencies. LPF: Low-pass filter.
(c) The beam splitter gate is conditioned on the internal spin state of the ion, spanned by $\ket{^2S_{1/2},F=0,m_F=0} = \ket{g}$, and the first order magnetically sensitive state $\ket{^2S_{1/2},F=1,m_F=1}=\ket{e}$.
(d) A running optical lattice induces a state dependent dipole force $F(t)$ that is modulated with $\omega_L$. The CBS Hamiltonian (\ref{eq:CBS_H}) is realized when $\omega_L = |\omega_\textrm{x}-\omega_\textrm{y}|$.
}
\end{figure}

By preparing the motional modes in the Fock states $\ket{0}$ or $\ket{1}$ and applying the conditional beam splitter gate, one obtains a Fredkin gate (Fig.~\ref{fig:fredkin_table}a), where the state of the motional modes are swapped only if the spin is in the excited state $\ket{e}$~\cite{Fredkin1982,Patel2016, Ono2017,Linke2018,Gao2019,Zhang2019}. The Fock state $\ket{1}$ is prepared starting from the vacuum state $\ket{0}$ by applying a $\pi$-pulse to the blue sideband of the corresponding motional mode~\cite{Meekhof1996,Wineland1998}. We prepare all 8 possible basis states (See Fig. \ref{fig:fredkin_table}) and measure the probability of finding the system in one of the states after applying the CBS transformation (Eq.~\ref{eq:CBS_U}). 

We measure the projection onto basis states by 3 consecutive projective measurements onto the spin state $\ket{g}$. We first perform state detection on the spin. If fluorescence is not detected, the internal state is projected into $\ket{g}$ and the motional state does not change. Then, we apply the red sideband (rsb) $\pi$-pulse to the mode $a$, followed by another spin detection. This projects the ion into the state $\ket{g,0}$ if fluorescence is not detected. The same operation is repeated for mode $b$ to find the probability of the ion in state $\ket{g,0,0}$. The measurement sequence for other basis states are similar. For example, to measure $\ket{g,1,0}$ we apply a $\pi$-pulse on the carrier transition before the first rsb $\pi$-pulse. 

The measured probability of obtaining each output basis state is shown in Fig.~\ref{fig:fredkin_table}b, which agrees with the expected behavior of a Fredkin gate.
\begin{figure}
\centering
\includegraphics[height=0.8\columnwidth]{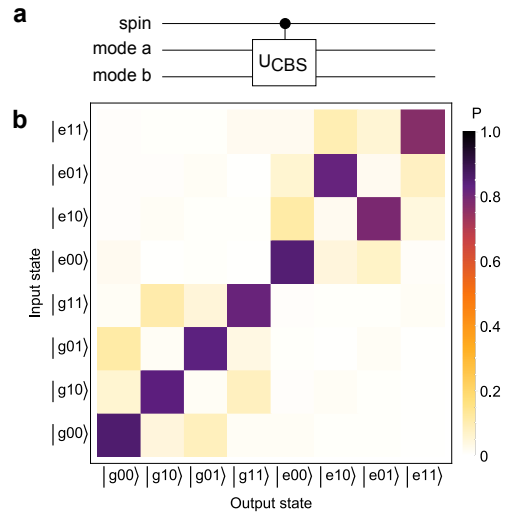}
\caption{
\label{fig:fredkin_table}
Fredkin gate truth table. By limiting the two radial modes of the ion to a single phonon each, the conditional beam splitter is able to simulate a quantum Fredkin gate. (a) Quantum circuit for the Fredkin gate. (b) For each of the eight basis states prepared as input, the probabilities of all outcomes after applying the CBS gate are measured via a series of projective measurements. Each data square represents an average of 10000 experiments. An average gate success probability of $0.82\pm\,0.01$ is obtained.
}
\end{figure}
Without correcting for any state preparation and measurement (SPAM) errors, we obtain a gate success probability~\cite{Patel2016,Ono2017,Linke2018} of $0.82\pm\,0.01$.

Generalization of the Fredkin gate to states $\ket{\psi},\ket{\phi}$ in the Hilbert space of a larger dimension is the controlled-swap (CSWAP) gate, which applies the transformation $\mathcal{CS}\ket{e,\psi,\phi}=\ket{e,\phi,\psi}$ if the control qubit is in state $\ket{e}$, and does not change the state if the control qubit is in state $\ket{g}$. The CSWAP gate has a number of known applications, including purity measurement, wavefunction overlap~\cite{Radim2002}, quantum fingerprinting~\cite{Buhrman2001}, quantum online memory checking~\cite{VanDam2010}, quantum digital signature~\cite{Gottesman2001}, quantum zero knowledge proof~\cite{Kobayashi2007}, and variational quantum algorithm~\cite{Preskill2018,Endo2018}. It is obvious from Eq.~(\ref{eq:CBS_U}) that the CBS and CSWAP gates are different due to the extra state dependent phase factor. However, for input states which are eigenstates of the parity operator $\hat{P}$ both gates produce similar results.

An example is the measurement of overlap between two states, $\braket{\phi|\psi}$, also known as a swap test~\cite{Buhrman2001,Gottesman2001}. With an initial state $\ket{g,\psi,\phi}$, applying the circuit shown in Fig.~\ref{fig:overlap}a gives a probability $(1-|\braket{\phi|\psi}|^2)/2$ to observe the spin in the state $\ket{e}$. The overlap $|\braket{\phi|\psi}|^2$ between states can thus be found from measuring the spin.

For the CBS gate, the swap test is carried out between an arbitrary state $\ket{\psi}$ and a Fock state $\ket{m}$. We apply the sequence $\hat{R}(\frac{\pi}{2},\varphi)\hat{U}_\textrm{CBS}\hat{R}(\frac{\pi}{2},0)$ (Fig.~\ref{fig:overlap}b) to an initial state $\ket{g,\psi,m}$. This yields the probability to measure the state $\ket{e}$ at the end of the sequence:
\begin{equation}
\label{eq:swap_test}
P = \frac{1}{2} \left[ 1 - (-1)^{m+1}\cos(\varphi)\lvert \braket{m \vert \psi} \rvert^2 \right],
\end{equation}
where $\hat{R}(\theta,\varphi) = \left( \begin{smallmatrix} \cos \frac{\theta}{2} & -i\sin \frac{\theta}{2}\exp{(-i\varphi)} \\ -i\sin \frac{\theta}{2}\exp{(i\varphi)} & \cos \frac{\theta}{2} \end{smallmatrix} \right)$ is implemented by a microwave pulse with pulse area $\theta$ and phase $\varphi$, resonant with the $\ket{g}\rightarrow\ket{e}$ transition.

After the first $\pi/2$-pulse, the spin is in the superposition $(\ket{g}-i\ket{e})/\sqrt{2}$. In our experiment, the coherence time of the spin is $\approx$ 1.7\,ms, which is comparable to the duration of a single CBS gate ($\tau\approx400\,\mu$s). To alleviate the effects of spin decoherence (mainly due to magnetic field fluctuations), we integrate spin echo into the gate sequence. A single CBS gate is split into two, each with a duration $\tau/2$, and a microwave $\ket{g}\rightarrow\ket{e}$ $\pi$-pulse applied in between. The second CBS gate $\hat{U}(\tau/2)$ has a phase $\upsilon = \pi$ relative to the first. Implementing this spin echo technique increases the spin coherence time to $\sim7$\,ms. We note that applying the spin echo does not preserve the transformation Eq.~(\ref{eq:CBS_U}) exactly, but the outcomes of the algorithms remain unchanged~\cite{supplemental}.

We begin by measuring the overlap between Fock states $\ket{\psi,\phi}=\ket{n,m}$, for $n,m \in [0,5]$. 
By varying the phase $\varphi$ of the last microwave $\pi/2$-pulse with respect to the first from 0 to $2\pi$, we obtain the oscillating probability (Eq.~\ref{eq:swap_test}) to observe $\ket{e}$ with amplitude equal to the overlap. We perform 300 experiments for each 24 steps of the phase $\varphi$.
The results shown in Fig.~\ref{fig:overlap}b are in agreement with the expected outcome of highest overlap probability along the diagonal. 
At higher Fock states we see reduction of overlap along the diagonal elements, and increasing overlap between $n$ and $m=n\pm1$. We attribute this to heating of the motional modes that increases with $n$.
\begin{figure}
\centering
\includegraphics[width=\columnwidth]{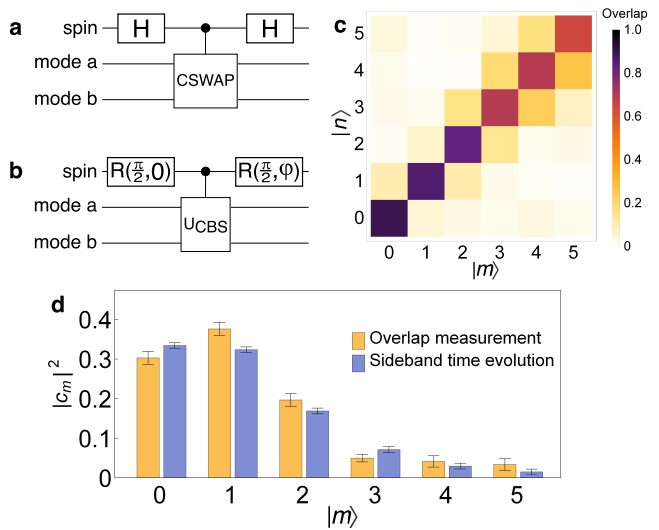}
\caption{
\label{fig:overlap}
Quantum circuit for the swap test with (a) CSWAP gate, and (b) CBS gate. H: Hadamard gate, $\hat{R}(\frac{\pi}{2},\varphi)$: $\pi/2$ spin rotation pulses with phase $\varphi$. (c) Overlap measurement between phonon modes, constrained to Fock states $\ket{n}$ and $\ket{m}$. Varying the phase $\varphi$ from 0 to $2\pi$ gives rise to an oscillating probability to observe $\ket{e}$. 300 experiments for each 24 step of $\varphi$ were performed, with each square representing data extracted from the contrast. (d) A coherent state prepared in one mode can be reconstructed by measuring overlap with Fock states prepared in the other. Using the same overlap measurement method as per (c) yielded $|\alpha|^2=1.9(2)$, compared to $|\alpha|^2=1.8(1)$ obtained from Fourier analysis of the spin state time evolution while driving the blue motional sideband. For the overlap measurement, 500 experiments for each 24 step of $\varphi$ were performed. Error bars denote one standard error of the mean (S.E.M.).
}
\end{figure}

An application of the overlap measurement is reconstruction of the phonon number population of a motional state. For an arbitrary state $\ket{\psi}=\sum_n c_n\ket{n}$, $|c_n|^2$ can be determined by measuring the overlap of $\ket{\psi}$ with $\ket{n}$. We demonstrate this by preparing a coherent state $\ket{\alpha}= \exp(-|\alpha|^2/2) \Sigma_n \alpha^n/\sqrt{n!} \ket{n}$ and a Fock state $\ket{n}$. We carry out the overlap measurements in the same manner as before, but with 500 experiments per step of $\varphi$. 
For comparison, we extract the phonon number distribution by driving the blue motional sideband transition and carrying out Fourier analysis of the spin state time evolution ~\cite{Meekhof1996,Leibfried1996,Leibfried2003}. The overlap measurement yields $|\alpha|^2=1.9(2)$, while the Fourier transform method gives $|\alpha|^2=1.8(1)$ (Fig. \ref{fig:overlap}c). 

The parity of the state $\ket{\psi}$ can be measured by applying the gate sequence $\hat{R}(\frac{\pi}{2},0)\hat{U}^{2}_{CBS}\hat{R}(\frac{\pi}{2},0)$ as shown in Fig.~\ref{fig:wigner_measurement}a, to the initial state $\ket{g,\psi,0}$ which produces the final state
\[
\Bigg[
\sum_{n=\textrm{odd}} c_n \ket{g}\ket{n} - i\sum_{n=\textrm{even}}c_n\ket{e}\ket{n} \Bigg] \otimes \ket{0}.
\]
Measuring the spin thus provides information about parity, which enables reconstruction of the Wigner function $W(\alpha)$~\cite{Royer1977,Bertet2002}. The Wigner function can be written as $W(\alpha)=\frac{2}{\pi}Tr[\hat{D}(-\alpha)\rho \hat{D}(\alpha) \hat{P}]$, where $\hat{D}(\alpha)=\exp (\alpha\hat{a}^{\dagger}-\alpha^*\hat{a})$ is the displacement operator, and $\rho$ is a density matrix. Therefore the value of $W(\alpha)$ can be obtained by measuring the parity of the state after it has been displaced by $\alpha$ in the phase space. We perform the displacement by applying the optical dipole force modulated at the frequency of the corresponding mode (Fig.~\ref{fig:exp_setup}d) with controlled phase and duration~\cite{Ding2014}. 

The Wigner function of a Fock state $\ket{n}$ is $W_n(\alpha)=\frac{2}{\pi}(-1)^n \exp(-2|\alpha|^2)L_n(4|\alpha|^2)$ , where $L_n$ is the Laguerre polynomial~\cite{BarnettRadmore2002}.
\begin{figure}
\centering
\includegraphics[width=\columnwidth]{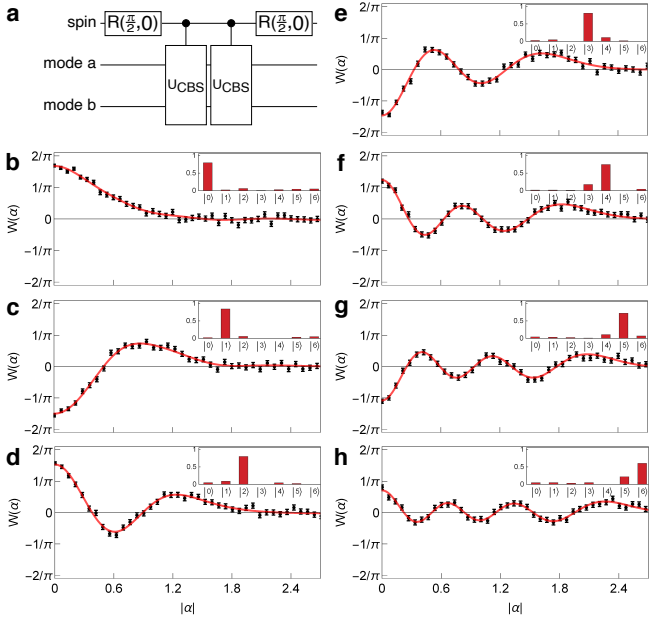}
\caption{
\label{fig:wigner_measurement}
Wigner function measurement. (a) Quantum circuit for single-shot parity measurement. (b-h) Wigner functions for Fock states from $n=0$ to $6$. The solid line fit to the data points assume a superposition of Fock states from $n=0$ to $6$, and each data point consists of 600 experimental runs. The error bars denote the S.E.M. (Insets) Phonon number distribution obtained from the fit.
}
\end{figure}
Measurement results of $W_n({\alpha})$ for modes initially prepared in Fock states $n=0$ to $6$ are shown in Fig.~\ref{fig:wigner_measurement}(b-h). Due to imperfection in state preparation and heating effects, the experimental data deviates from the ideal case of a pure Fock state with increasing $n$. To account for these imperfections and to determine the phonon number distribution, the data are fit to a linear superposition of Wigner functions $\sum_n d_n W_n(\alpha)$, as the form of the density matrix is expected to be $\rho=\sum_n d_n\ket{n}\bra{n}$. The fit is shown as a solid line in Fig.~\ref{fig:wigner_measurement}(b-h). The effect of heating grows with increasing $n$ as shown from the inset in Fig. \ref{fig:wigner_measurement}.

The CBS gate enables generation of NOON states $\ket{\psi_\textrm{NOON}}=(\ket{n,0}+\ket{0,n})/\sqrt{2}$, similar to the method proposed in \cite{HKLau2016,HKLau2017,Buzek2000}. Despite its usefulness in quantum metrology~\cite{Sanders1989,Kok2002}, few experiments have demonstrated NOON state preparation for $n>2$~\cite{Mitchell2004,Nagata2007,Afek2010,Zhang2018}.

We implement a deterministic NOON state preparation algorithm with a constant circuit depth (Fig.~\ref{fig:NOON_Fidelity_Fisher}a) that is independent of $n$. Starting with an initial state $\ket{g,n,0}$, the sequence
\[
\hat{R}(\frac{\pi}{2},0) \hat{U}_\textrm{CBS} \hat{R}(\frac{\pi}{2},\varphi) \hat{U}_\textrm{CBS} \hat{R}(\frac{\pi}{2},0)
\]
generates a NOON state. The phase factor introduced by the CBS gate imposes a number dependence: when $n$ is odd, we require $\varphi=0$, while for even $n$, $\varphi = \pi/2$.
\begin{figure}
\centering
\includegraphics[width=\columnwidth]{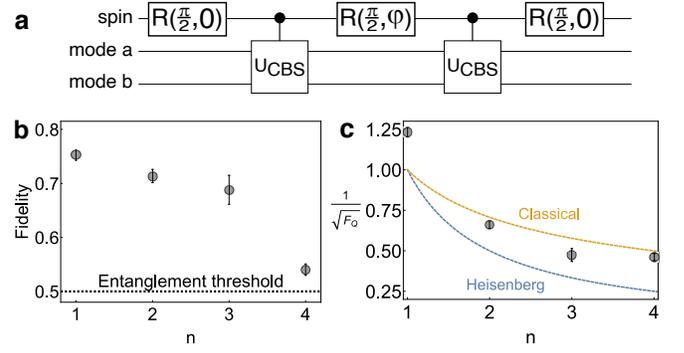}
\caption{
\label{fig:NOON_Fidelity_Fisher}
Generation of NOON states. (a) Quantum circuit for NOON state generation. The quantum circuit is parity dependent: $n$ = odd requires $\varphi=0$, while $n$ = even requires $\varphi=\pi/2$. (b) Fidelity of the prepared NOON state as a function of $n$. Fidelity above 0.5 implies that the state is entangled. (c) Quantum Fisher information of the prepared NOON states. Apart from $n = 1$, the prepared states violate the classical lower bound. Error bars are propagated from measurements of the matrix element of $\rho_\textrm{exp}$.
}
\end{figure}
Figure \ref{fig:NOON_Fidelity_Fisher} shows the fidelity $\mathcal{F} = \bra{\psi_\textrm{NOON}}\rho_\textrm{exp}\ket{\psi_\textrm{NOON}} = (P_{n,0}+P_{0,n}+2\rho_{n0,0n})/2$ and quantum Fisher information $F_Q=n^2 (2\rho_{n0,0n})^2/(P_{n,0}+P_{0,n})$~\cite{Zhang2018} for the prepared NOON states from $n=1$ to $4$~\cite{supplemental}. Here $P_{n,0} = \bra{n,0} \rho_\textrm{exp} \ket{n,0}$ and $\rho_{n0,0n} = \bra{n,0} \rho_\textrm{exp} \ket{0,n} = \bra{0,n} \rho_\textrm{exp} \ket{n,0}$.
An ideal NOON state offers optimal scaling of phase sensitivity $\Delta \phi$ that approaches the Heisenberg limit $\Delta \phi = 1/n$.
We see that our prepared states allows measurement sensitivity that is only slightly better than that offered by states based on classical correlations. This deviation from the ideal case is attributed to decoherence of the motional state due to fast fluctuations of the radial trap frequencies. For the phonon superposition $(\ket{0}+\ket{1})/\sqrt{2}$, the $x$ and $y$ modes coherence times were measured to be 5.0(7) and 7(1)ms, respectively. However, for the superposition $(\ket{0}+\ket{2})/\sqrt{2}$, the coherence time decreases to 1.2(3)ms and 1.4(3)ms. As the decoherence rate scales like $\lvert n_a - n_b \rvert^2$~\cite{Myatt2000}, at higher $n$ the superposition $(\ket{0}+\ket{n})/\sqrt{2}$ rapidly dephases during the generation sequence. Improving the coherence time would improve the fidelity of the generated NOON states with larger $n$. 

The combination of CBS gate and parity operator removes the state-dependent phase factor in Eq.~\ref{eq:CBS_U}, and implements a CSWAP gate~\cite{Radim2002,HKLau2016}. This requires an additional ancillary mode $c$ initially prepared in vacuum state to implement a parity operator, and along with astute use of the phase $\upsilon$ in Eq.~\ref{eq:CBS_H}, can be realized as~\cite{Wang2001,Radim2002}:
\begin{equation}
\label{eq:CBS_CSWAP}
\mathcal{CS} = \hat{U}^{ab,~\upsilon=\pi/2}_\textrm{CBS}  (\hat{U}^{ac,~\upsilon=0}_\textrm{CBS})^2,
\end{equation}
where the letters in the superscript denote modes that the gates are applied to. The beam splitter coupling between $a,c$ and $a,b$ is realized in the same manner. This enables implementation of an exponential swap (ESWAP) gate~\cite{HKLau2016,HKLau2017}, which can be used for deterministic entanglement generation~\cite{Buzek2000}, quantum principal component analysis (QPCA), and matrix inversion algorithms~\cite{Biamonte2017,Harrow2009,Lloyd2014,HKLau2017}. The gate may also find uses in quantum thermodynamics~\cite{Alicki2018}, quantum metrology and sensing~\cite{Mitchell2004}.

\begin{acknowledgements}
This research is supported by the National Research Foundation, Prime Ministers Office, Singapore, and the Ministry of Education, Singapore, under the Research Centers of Excellence program and Education Academic Research Fund Tier 2 (Grant No. MOE2016-T2-1-141).
\end{acknowledgements}

\setcounter{equation}{0}
\setcounter{figure}{0}
\setcounter{table}{0}
\makeatletter
\renewcommand{\theequation}{S\arabic{equation}}
\renewcommand{\thefigure}{S\arabic{figure}}


%
\begin{widetext}
\begin{center}
\textbf{\large Supplemental materials: Hybrid quantum computation gate with trapped ion system}
\end{center}

\date{\today}
\maketitle

\section{Effects of spin echo on measurements}
In the experiments, we incorporate spin echo pulses into sequences in order to improve coherence time of the spin. This modifies the state evolution, but does not change the outcome of the algorithms. Analysis of each algorithm is carried out below.

For the beam splitter Hamiltonian 
\[
\hat{H}_\textrm{BS} = \hbar \xi \left( \hat{a}^\dagger \hat{b} \operatorname{e}^{i\upsilon} + \hat{a} \hat{b}^\dagger \operatorname{e}^{-i\upsilon} \right),
\]
we denote its time evolution operation as $\hat{U}_\textrm{BS}(t,\upsilon) = \exp \left(-\frac{i}{\hbar}\hat{H}_\textrm{BS}t\right)$. The conditional beam splitter Hamiltonian is $\hat{H}_\textrm{CBS} = \hat{H}_\textrm{BS} \otimes \ket{e}\bra{e}$, and its time evolution operator is $\hat{U}(t,\upsilon) = \hat{U}_\textrm{BS}(t,\upsilon) \otimes \ket{e}\bra{e} + \hat{I}\otimes\ket{g}\bra{g}$. Notation of the states are always ordered in the manner of spin, mode $a$, and mode $b$.

\subsection{Swap test}
The algorithm for swap test on an intial state $\ket{g,\psi,m}$ with the conditional beam splitter is
\begin{align}
\label{eq:SMswaptest}
\hat{R}\left(\frac{\pi}{2},0\right) \hat{U}\left(\frac{\pi}{2\xi},0\right) \hat{R}\left(\frac{\pi}{2},0\right) \ket{g,\psi,m},
\end{align}
Evaluating the operations in Eq.~\ref{eq:SMswaptest} gives the final state
\begin{align}
\ket{\Psi} = \frac{1}{2} \left[ \ket{g}\left( \hat{I} - \hat{U}_\textrm{BS}\left(\frac{\pi}{2\xi},0\right) \right) \ket{\psi,m}  -i\ket{e} \left( \hat{I} + \hat{U}_\textrm{BS}\left(\frac{\pi}{2\xi},0\right) \right)\ket{\psi,m} \right].
\end{align}
The overlap is found from the probability to detect the spin in state $\ket{g}$, which we evaluate to get
\begin{align}
\label{eq:SMswaptestmeasure}
\operatorname{tr} \left[ \ket{g}\braket{g\vert \Psi} \bra{\Psi} \right] = \frac{1}{2} - \frac{1}{4} \left[ \bra{\psi,m} \hat{U}_\textrm{BS}^\dagger\left(\frac{\pi}{2\xi},0\right)\ket{\psi,m} + \bra{\psi,m}\hat{U}_\textrm{BS}\left(\frac{\pi}{2\xi},0\right)\ket{\psi,m}\right].
\end{align}
With spin echo, the entire sequence becomes
\begin{align}
\label{eq:SMswaptestspinecho}
\hat{R}\left(\frac{\pi}{2},0\right) \hat{U}\left(\frac{\pi}{4\xi},\pi\right) \hat{R}\left(\pi,0\right) \hat{U}\left(\frac{\pi}{4\xi},0\right) \hat{R}\left(\frac{\pi}{2},0\right) \ket{g,\psi,m},
\end{align}
which results in the final state
\begin{align}
\ket{\Psi^\prime} = \frac{1}{2} \left[ -\ket{g}\left( \hat{U}_\textrm{BS}\left(\frac{\pi}{4\xi},0\right) + \hat{U}_\textrm{BS}\left(\frac{\pi}{4\xi},\pi\right)\right)\ket{\psi,m} +i\ket{e} \left( \hat{U}_\textrm{BS}\left( \frac{\pi}{4\pi},0\right) - \hat{U}_\textrm{BS}\left(\frac{\pi}{4\xi},\pi\right) \right)\ket{\psi,m} \right].
\end{align}
Working out the probability to detect the spin in the state $\ket{e}$, we get
\begin{align}
\label{eq:SMswaptestmeasurespinecho}
\operatorname{tr} \left[ \ket{e} \braket{e\vert\Psi^\prime} \bra{\Psi^\prime} \right] = & \frac{1}{2} - \frac{1}{4} \bra{\psi,m} \hat{U}_\textrm{BS}^\dagger\left(\frac{\pi}{4\xi},0\right) \hat{U}_\textrm{BS}\left(\frac{\pi}{4\xi},\pi\right) + \hat{U}_\textrm{BS}^\dagger\left(\frac{\pi}{4\xi},\pi\right)\hat{U}_\textrm{BS}\left(\frac{\pi}{4\xi},0\right) \ket{\psi,m} \nonumber \\
= & \frac{1}{2} - \frac{1}{4} \left[ \bra{\psi,m} \hat{U}_\textrm{BS}^\dagger \left(\frac{\pi}{2\xi},0\right)\ket{\psi,m} + \bra{\psi,m}\hat{U}_\textrm{BS}\left(\frac{\pi}{2\xi},0\right)\ket{\psi,m} \right],
\end{align}
where we have used the properties $\hat{U}_\textrm{BS}\left(t,\pi\right) = \hat{U}_\textrm{BS}^\dagger\left(t,0\right)$ and $\hat{U}_\textrm{BS}\left(t,0\right)\hat{U}_\textrm{BS}\left(t,0\right) = \hat{U}_\textrm{BS}\left(2t,0\right)$ to obtain the last line. We see that Eq.~\ref{eq:SMswaptestmeasurespinecho} is equivalent to Eq.~\ref{eq:SMswaptestmeasure}. The swap test therefore still works with spin echo; instead of detecting $\ket{g}$, one has to measure $\ket{e}$.

\subsection{Single shot parity measurement}
For the parity gate, we apply the following sequence to an initial state $\ket{g,\psi,0}$
\begin{align*}
\hat{R}\left(\frac{\pi}{2},0\right) \hat{U}\left(\frac{\pi}{\xi},0\right) \hat{R}\left(\frac{\pi}{2},0\right) \ket{g,\psi,0}.
\end{align*}
The final state after the sequence is
\begin{align*}
\ket{\Psi} = \frac{1}{2} \left[ \ket{g} \left( \hat{I} - \hat{U}_\textrm{BS}\left(\frac{\pi}{\xi},0\right) \right) \ket{\psi,0} - i \ket{e}\left( \hat{I} + \hat{U}_\textrm{BS} \left(\frac{\pi}{\xi},0\right) \right)\ket{\psi,0} \right]
\end{align*}
The outcome of spin measurements corresponds to parity of the state $\ket{\psi}$; for odd (even) parity the spin state $\ket{g}\,(\ket{e})$ is detected. Evaluating the outcomes, we get
\begin{align}
\label{eq:SMparitygateG}
\operatorname{tr} \left[ \ket{g}\braket{g\vert\Psi}\bra{\Psi} \right] & = \frac{1}{2} - \frac{1}{4} \bra{\psi,0} \left( \hat{U}_\textrm{BS}\left(\frac{\pi}{\xi},0\right) + \hat{U}_\textrm{BS}^\dagger\left(\frac{\pi}{\xi},0\right)\right) \ket{\psi,0}, \\
\label{eq:SMparitygateE}
\operatorname{tr} \left[ \ket{e}\braket{e\vert\Psi}\bra{\Psi} \right] & = \frac{1}{2} + \frac{1}{4} \bra{\psi,0} \left( \hat{U}_\textrm{BS}\left(\frac{\pi}{\xi},0\right) + \hat{U}_\textrm{BS}^\dagger\left(\frac{\pi}{\xi},0\right)\right) \ket{\psi,0}.
\end{align}
With spin echo, the modified sequence is
\begin{align*}
\hat{R}\left(\frac{\pi}{2},0\right) \hat{U}\left(\frac{\pi}{2\xi},\pi\right) \hat{R}\left(\pi,0\right)  \hat{U}\left(\frac{\pi}{2\xi},0\right) \hat{R}\left(\frac{\pi}{2},0\right) \ket{g,\psi,0}.
\end{align*}
This gives the final state
\begin{align*}
\ket{\Psi^\prime} = \frac{1}{2} \left[ -\ket{g} \left(\hat{U}_\textrm{BS}\left(\frac{\pi}{2\xi},0\right) + \hat{U}_\textrm{BS}\left(\frac{\pi}{2\xi},\pi\right) \right)\ket{\psi,0} + i\ket{e}\left(\hat{U}_\textrm{BS}\left(\frac{\pi}{2\xi},0\right) - \hat{U}_\textrm{BS}\left(\frac{\pi}{2\xi},\pi\right)\right)\ket{\psi,0}\right].
\end{align*}
Evaluating the spin measurement outcomes,
\begin{align}
\label{eq:SMparitygatespinechoG}
\operatorname{tr} \left[ \ket{g}\braket{g\vert\Psi^\prime}\bra{\Psi^\prime} \right] & = \frac{1}{2} + \frac{1}{4} \bra{\psi,0} \left( \hat{U}_\textrm{BS}^\dagger \left(\frac{\pi}{2\xi},0\right) \hat{U}_\textrm{BS}\left(\frac{\pi}{2\xi},\pi\right) + \hat{U}_\textrm{BS}^\dagger\left(\frac{\pi}{2\xi},\pi\right)\hat{U}_\textrm{BS}\left(\frac{\pi}{2\xi},0\right)\right)\ket{\psi,0} \nonumber \\
& = \frac{1}{2} + \frac{1}{4} \bra{\psi,0} \left( \hat{U}_\textrm{BS}^\dagger\left(\frac{\pi}{\xi},0\right) + \hat{U}_\textrm{BS}\left(\frac{\pi}{\xi},0\right) \right)\ket{\psi,0} \\
\label{eq:SMparitygatespinechoE}
\operatorname{tr} \left[ \ket{e}\braket{e\vert\Psi^\prime}\bra{\Psi^\prime} \right] & = \frac{1}{2} - \frac{1}{4} \bra{\psi,0} \left( \hat{U}_\textrm{BS}^\dagger \left(\frac{\pi}{2\xi},0\right) \hat{U}_\textrm{BS}\left(\frac{\pi}{2\xi},\pi\right) + \hat{U}_\textrm{BS}^\dagger\left(\frac{\pi}{2\xi},\pi\right)\hat{U}_\textrm{BS}\left(\frac{\pi}{2\xi},0\right)\right)\ket{\psi,0} \nonumber \\
& = \frac{1}{2} - \frac{1}{4} \bra{\psi,0} \left( \hat{U}_\textrm{BS}^\dagger\left(\frac{\pi}{\xi},0\right) + \hat{U}_\textrm{BS}\left(\frac{\pi}{\xi},0\right) \right)\ket{\psi,0}
\end{align}
Comparison between the outcomes (Equations \ref{eq:SMparitygateG}, \ref{eq:SMparitygateE}, \ref{eq:SMparitygatespinechoG}, and \ref{eq:SMparitygatespinechoE}) show that implementing the spin echo pulse still allows parity to be measured, with the only difference being the associated parity and spin states. i.e. with spin echo, odd (even) parity corresponds to the spin state $\ket{e}\,(\ket{g})$.

\begin{figure*}[!h]
    \centering
    \includegraphics[width=0.55\columnwidth]{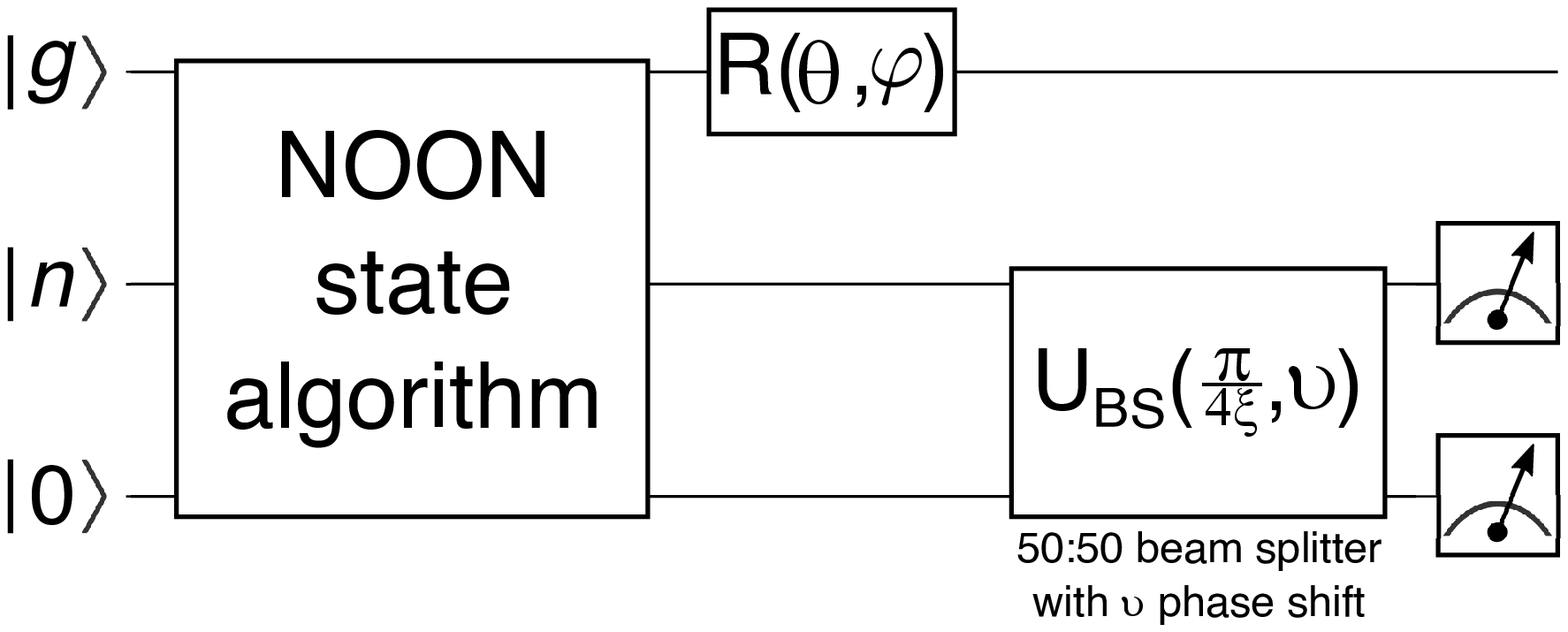}
    \caption{Circuit diagram for measuring off-diagonal density matrix elements. The NOON state is first produced by applying the NOON generation algorithm. A spin rotation pulse $\hat{R}(\theta,\phi)$ is then applied, where $\theta$ and $\phi$ are chosen such that the ensuing spin state is $\ket{e}$. Applying the running optical lattice with phase $\upsilon$ for a duration $\frac{\pi}{4\xi}$ gives rise to a 50:50 beam splitter transformation. The parity of each mode is measured by reconstructing its phonon number distribution, and is repeated for varying 50:50 beam splitter phase $\upsilon$.} 
    \label{fig:SMNOONoffdiagcircuit}
\end{figure*}

\begin{figure*}[!h]
    \centering
    \includegraphics[width=0.8\columnwidth]{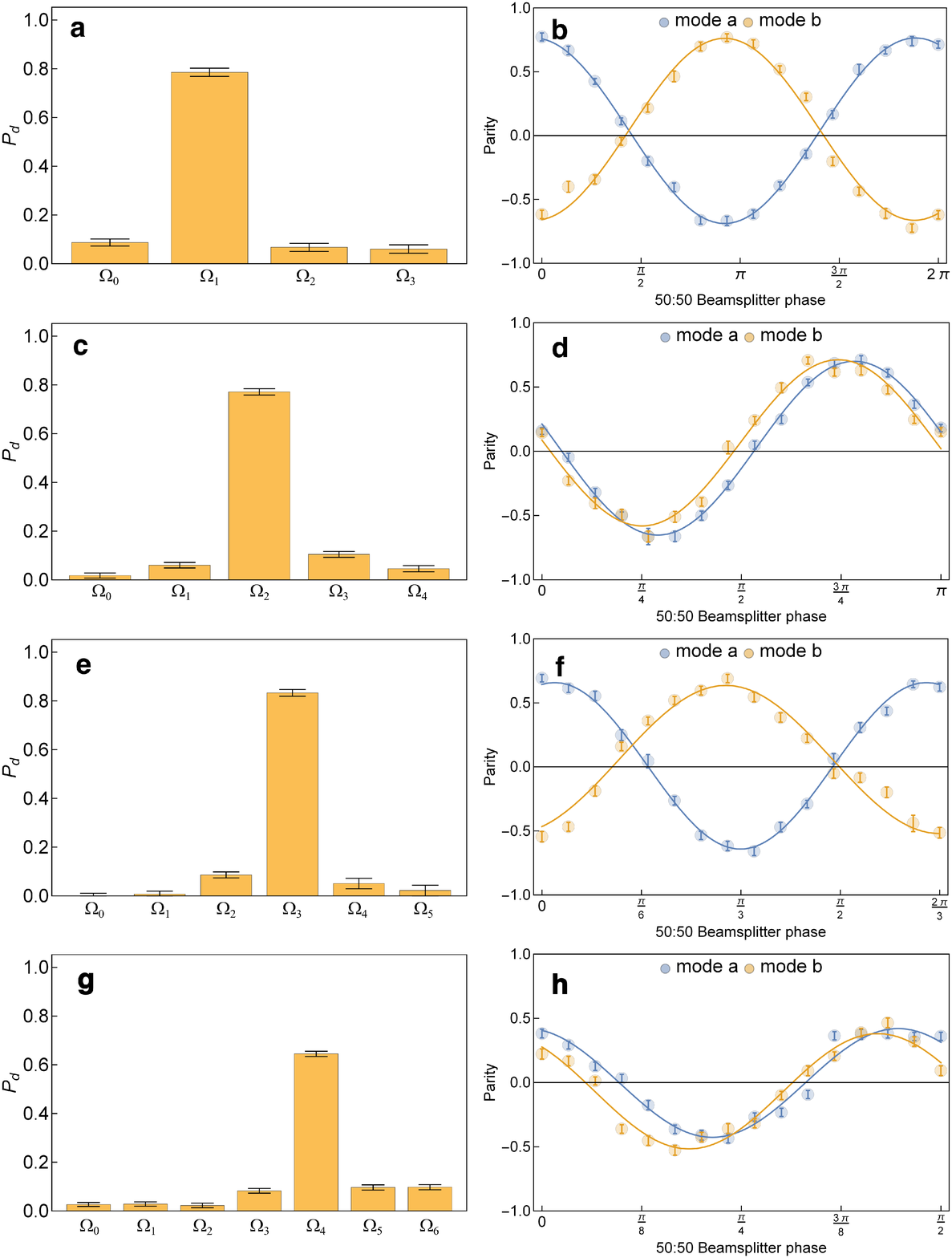}
    \caption{Density matrix element measurements of experimentally prepared NOON states from (a,b) $n=1$ to (g,h) $n=4$. 
    (Left) By driving the ``joint'' blue motional sideband $\hat{H}_\textrm{joint}$ (Eq.~\ref{eq:SMjointbsb}), diagonal elements can be measured from Fourier transforming the time evolution of the probability to detect the spin in the excited state. The extracted population $P_d$ for each component $\Omega_d$ is shown.
    (Right) Measurements of the off-diagonal elements are carried out by first subjecting the prepared NOON states to a phase shift of $\upsilon$ (Fig.~\ref{fig:SMNOONoffdiagcircuit}), followed by a 50:50 beam splitter operation. This is done by applying the running optical lattice with a phase $\upsilon$. The parity of each mode is then measured via reconstruction of the phonon number distribution. Shown in the figure are the measured parity versus beam splitter phase, where the contrast of parity oscillation yields the sum of off-diagonal elements for each $n$.} 
    \label{fig:SMNOON}
\end{figure*}

\subsection{NOON state generation}
To generate NOON states, the algorithms without spin echo are
\begin{align*}
n:\textrm{odd} & \Rightarrow \hat{R}\left(\frac{\pi}{2},0\right) \hat{U}\left(\frac{\pi}{2\xi},0\right) \hat{R}\left(\frac{\pi}{2},0\right) \hat{U}\left(\frac{\pi}{2\xi},0\right) \hat{R}\left(\frac{\pi}{2},0\right) \ket{g,n,0} \\
n:\textrm{even} & \Rightarrow  \hat{R}\left(\frac{\pi}{2},0\right) \hat{U}\left(\frac{\pi}{2\xi},0\right) \hat{R}\left(\frac{\pi}{2},\frac{\pi}{2}\right) \hat{U}\left(\frac{\pi}{2\xi},0\right) \hat{R}\left(\frac{\pi}{2},0\right) \ket{g,n,0} \\
\end{align*}
For odd $n$, the final state obtained is
\begin{align*}
\ket{\Psi_\textrm{odd}} & = \frac{1}{2\sqrt{2}} \left[ \ket{g} \left( \hat{I} - 2\hat{U}_\textrm{BS}\left(\frac{\pi}{2\xi},0\right) - \hat{U}_\textrm{BS}\left(\frac{\pi}{\xi},0\right) \right) \ket{n,0} - i\ket{e}\left( \hat{I} + \hat{U}_\textrm{BS}\left(\frac{\pi}{\xi},0\right) \right) \ket{n,0} \right] \\
& = \frac{1}{\sqrt{2}} \ket{g} \left(\ket{n,0} - \left(-i\right)^n \ket{0,n}\right),
\end{align*}
where the result $\hat{U}_\textrm{BS}\left(\frac{\pi}{\xi},0\right)\ket{n,0} = -\ket{n,0}$ for odd $n$ was used to get the last line. Similarly for even $n$, we get
\begin{align*}
\ket{\Psi_\textrm{even}} & = \frac{1}{2\sqrt{2}} \left[ \ket{g} \left(  \hat{I} - \hat{U}_\textrm{BS}\left(\frac{\pi}{\xi},0\right) \right) \ket{n,0} + \ket{e}\left(2\hat{U}_\textrm{BS}\left(\frac{\pi}{2\xi},0\right) - i\left( \hat{I} + \hat{U}_\textrm{BS}\left(\frac{\pi}{\xi},0\right) \right) \right) \ket{n,0} \right] \\
& = \frac{1}{\sqrt{2}} \ket{e}\left(-i\ket{n,0} + \left(-i\right)^n\ket{0,n}\right),
\end{align*}
where the result $\hat{U}_\textrm{BS}\left(\frac{\pi}{\xi},0\right)\ket{n,0} = \ket{n,0}$ for even $n$ was used. 

With the inclusion of spin echo, the algorithms are
\begin{align*}
n:\textrm{odd} & \Rightarrow \hat{R}(\frac{\pi}{2},0) \hat{U}(\frac{\pi}{4\xi},\pi) \hat{R}(\pi,0) \hat{U}(\frac{\pi}{4\xi},0) \hat{R}(\frac{\pi}{2},0) \hat{U}(\frac{\pi}{4\xi},\pi) \hat{R}(\pi,0) \hat{U}(\frac{\pi}{4\xi},0) \hat{R}(\frac{\pi}{2},0) \ket{g,n,0} \\
n:\textrm{even} & \Rightarrow \hat{R}(\frac{\pi}{2},0) \hat{U}(\frac{\pi}{4\xi},\pi) \hat{R}(\pi,0) \hat{U}(\frac{\pi}{4\xi},0) \hat{R}(\frac{\pi}{2},\frac{\pi}{2}) \hat{U}(\frac{\pi}{4\xi},\pi) \hat{R}(\pi,0) \hat{U}(\frac{\pi}{4\xi},0) \hat{R}(\frac{\pi}{2},0) \ket{g,n,0}
\end{align*}
For odd $n$, we get the final state
\begin{align*}
\ket{\Psi^\prime_\textrm{odd}} & = \frac{1}{2\sqrt{2}} \left[ \ket{g} \left( \hat{U}_\textrm{BS}\left(\frac{\pi}{2\xi},0\right) + \hat{U}_\textrm{BS}\left(\frac{\pi}{2\xi},\pi\right) \right) \ket{n,0} + i\ket{e} \left( 2\hat{I} + \hat{U}_\textrm{BS}\left(\frac{\pi}{2\xi},\pi\right) - \hat{U}_\textrm{BS}\left(\frac{\pi}{2\xi},0\right) \right) \ket{n,0} \right] \\
& = \frac{1}{\sqrt{2}} i\ket{e} \left(\ket{n,0} - \left(-i\right)^n \ket{0,n} \right),
\end{align*}
and for even $n$,
\begin{align*}
\ket{\Psi^\prime_\textrm{even}} & = \frac{1}{2\sqrt{2}} \left[ i\ket{g} \left( \hat{U}_\textrm{BS}(\frac{\pi}{2\xi},0) - \hat{U}_\textrm{BS}(\frac{\pi}{2\xi},\pi) \right) \ket{n,0} + \ket{e} \left( 2i\hat{I} + \hat{U}_\textrm{BS}(\frac{\pi}{2\xi},0) + \hat{U}_\textrm{BS}(\frac{\pi}{2\xi},\pi) \right) \ket{n,0} \right] \\
& = \frac{1}{\sqrt{2}} \ket{e} (i\ket{n,0} - (-i)^n \ket{0,n} ).
\end{align*}
The transformation $\hat{U}_\textrm{BS}\left(\frac{\pi}{2\xi},\theta\right)\ket{n,m} = \left(-i\right)^{n+m}\operatorname{e}^{i\left(m-n\right)\theta}\ket{m,n}$ was used for both odd and even cases to obtain the final state. For both odd and even $n$, NOON states can still be produced with the inclusion of spin echo pulses.

\section{NOON state analysis}
During the NOON state generation experiment the density matrix $\rho_\textrm{exp}$ is produced. To determine its fidelity $\mathcal{F} = \bra{\psi_\textrm{NOON}} \rho_\textrm{exp} \ket{\psi_\textrm{NOON}}$, we need information of both the diagonal and off-diagonal density matrix elements~\cite{Zhang2018}. Diagonal elements correspond to the population of the motional states $\ket{n,0}$ and $\ket{0,n}$, which we measure by making use of the ``joint'' blue sideband 
\begin{align}
\label{eq:SMjointbsb}
\hat{H}_\textrm{joint} = \hbar \Omega_d \left(\hat{a}^\dagger \hat{b}^\dagger \hat{\sigma}_+ + \hat{a}\hat{b}\hat{\sigma}_-\right).
\end{align}
Here $\hat{\sigma}_\pm$ refer to the usual spin raising and lowering operators. The Hamiltonian Eq.~\ref{eq:SMjointbsb} couples the states $\ket{n_a,n_b}\leftrightarrow\ket{n_a+1,n_b+1}$ with a Rabi frequency $\Omega_d =  \sqrt{d+1}\Omega_0 =  \sqrt{\left(n_a+1\right)\left(n_b+1\right)}\Omega_0$, where $d = n_a n_b + n_a + n_b $. Experimentally we implement this interaction by driving the second order motional sideband detuned by $\omega_a +\omega_b$ from the $\ket{{}^2S_{1/2},F=0,m_F=0}$ to $\ket{{}^2S_{1/2},F=1,m_F=0}$ transition. Time evolution of the probability to detect the spin in the excited state while driving the joint blue sideband can then be expressed as 
\begin{align}
P = \frac{1}{2} \left( 1 - \sum_{d=0}^\infty P_d \cos\left( \Omega_d t \right) \operatorname{e}^{-\gamma_d t} \right),
\end{align}
which allows extraction of the population $P_d$ by a Fourier transformation and determines population of the states $\ket{n,0}$ and $\ket{0,n}$. Results of extracting the Fourier components from time evolution measurement of the spin is shown in Fig.~\ref{fig:SMNOON}. For $n=1,2,4$, the components $P_d$ directly give the total population of the states $\ket{n,0}$ and $\ket{0,n}$. However for $n=3$, the states $\ket{3,0}$, $\ket{0,3}$, and $\ket{1,1}$ share the same Rabi frequency. To account for the contribution of the state $\ket{1,1}$, we additionally determine the population of the state $\ket{1}$ present in each of the $a$ and $b$ modes. The smaller value of the two gives an upper bound to the component of $\ket{1,1}$, and is deducted from $P_3$.

Measurement of the off-diagonal density matrix elements is done by subjecting the prepared NOON state to a phase shift $\upsilon$, followed by a 50:50 beam splitter. Experimentally we achieve this by varying the phase of the applied running lattice that gives rise to the 50:50 beam splitter. The parity of each mode is then measured by reconstructing the phonon number distribution (Fig.~\ref{fig:SMNOONoffdiagcircuit}). Oscillation of the parity as a function of phase shift indicates coherence, and its contrast gives the sum of off-diagonal matrix elements~\cite{Sackett2000}. The results of measuring parity versus phase of the applied 50:50 beam splitter is shown in Fig.~\ref{fig:SMNOON}.

\end{widetext}


\sloppy

\end{document}